\documentclass[a4paper,twocolumn,showpacs,preprintnumbers,amsmath,amssymb]{revtex4}

\usepackage{graphicx}
\usepackage{bm}
\newcommand{\dif}{\mathrm{d}}

\newcommand{\VV}{\mathbf{V}}
\newcommand{\PP}{\mathbf{P}}
\newcommand{\FF}{\mathbf{F}}
\newcommand{\vv}{\mathbf{v}}
\newcommand{\ww}{\mathbf{w}}
\newcommand{\II}{\mathbf{I}}

\begin{document}

\preprint{APS/123-QED}

\title{Force calculation on walls and
  embedded particles in \\
Multi-Particle Collision Dynamics simulations}
\author{A. Imperio$^1$} \email{a.imperio@virgilio.it} %
\textbackslash\textbackslash 

\author{J.T. Padding$^2$ and W. Briels$^1$}
 \homepage{http://cbp.tnw.utwente.nl/index.html} 
\affiliation{$^1$ Computational Biophysics, University of
Twente, P.O. Box 217, 7500 AE, The Netherlands\\
$^2$ IMCN, Universite catholique de Louvain, Croix du Sud 1,
Louvain-la-Neuve, Belgium
}%

\date{\today}

\begin{abstract} Colloidal solutions posses a wide range of time and
 length scales, so that it is unfeasible to keep track of all of
them within a single simulation. As a consequence some form of
coarse-graining must be
applied. In this work we use the Multi-Particle Collision Dynamics
scheme. We describe a particular implementation of no-slip
boundary conditions upon a solid surface, capable of providing correct forces on the solid bypassing the calculation of the velocity profile or the stress tensor in the fluid near the surface. As an application we measure the
friction on a spherical particle, when it is placed in a bulk fluid
 and when it is confined in a slit. We show that the
implementation of the no-slip boundary conditions leads to an enhanced Enskog friction, which can be understood analytically. 
Because of the long-range nature of hydrodynamic interactions, the
Stokes friction obtained from the simulations is sensitive of
 the simulation box size. We address this topic for the
slit geometry, showing that that the dependence on the system size
differs very much from what is expected in a 3D system, where periodic boundary
conditions are used in all directions.
\end{abstract}

\pacs{Valid PACS appear here}
\maketitle
\section{\label{I}Introduction}
The existence of a huge range of time and
 length-scales, spanning between mesoscopic colloidal
particles and microscopic solvent particles, constitutes a severe
 problem in numerical
simulations.
 Hybrid schemes have been developed, in which different coarse-grained approaches are used to describe solvent molecules and colloids.  Prominent in the class of solvent models is Multi-Particle
Collision Dynamics (MPCD), originally proposed
by Malevanets and Kapral \cite{malevanets99}, which has proved to be
very effective in simulating Newtonian fluids out of equilibrium. For a review see \cite{gompper09}.\\
The interaction between solvent molecules and colloids may be
 described in several ways, mimicking the various boundary conditions (BC) in use in continuum descriptions. The molecular
origins of these boundary conditions can be very complex
\cite{bocquet94,lauga05}, but
for most colloidal applications, simple no-slip BC are sufficient
for modeling experimental conditions. In this paper we discuss a way to impose such boundary conditions in MPCD simulations of liquids containing dissolved colloids.

MPCD simulations consist of two alternating steps: the streaming step
(where Newton's equations of motions for non-interacting particles are solved) and
the collision step (where the fluid is coarse-grained). The
no-slip BC involve both steps. In particular,
during the collision step, virtual particles (VP) are inserted
in the regions of the MPCD box occupied by walls or
colloids.This idea has already been used in  \cite{lamura01,winkler09,gotze07,downtown09}. In this paper,
we will provide additional insight in this method,
analyzing not only the velocity profiles nearby solid planar walls, but also
discussing the contributions of the VP to the forces exerted by the solvent on the
solid boundaries. Such forces determine the drag forces experienced by
the colloids and thereby regulate the colloids' dynamics. A recent
interesting work discussing these techniques in the implementation of
mixed stick-slip boundary conditions can be found in \cite{whitmer10}.\\
The second topic we treat is the role of finite-size effects (FSE) of
the simulation box. In molecular dynamics simulations it is common practice to
use periodic BC to mimic an infinite
system. However,  due to the long-range nature of the hydrodynamic
interactions,
quantities such as the Stokes friction can still be affected by the
system size even when equilibrium properties are well reproduced
\cite{dunweg93}.
 The use of periodic BC implies a set of images of the colloidal
 particle, so that they all together form a periodic grid.
The problem of slow flow through a 3D
periodic array of spheres has been treated by \cite{zick82},
while in \cite{dunweg93}
 FSE are studied for a chain of beads in a solvent. In both works,
 periodic BC are assumed along all three directions.\\
When confining walls are present in one direction, and
periodic BC in the others (such as in the simulation of a slit
geometry),  the problem of the FSE has been addressed only recently in \cite{kohale08} for
slip BC, using molecular dynamics simulations, and in \cite{bhattacharya08} for no-slip BC, using continuum
models. Our results, obtained with coarse-grained
simulations, are qualitatively in agreement with such previous works
for the  friction in the direction parallel to the
walls. Furthermore we also discuss the
friction in the direction perpendicular to the walls, which is missing
in \cite{bhattacharya08,kohale08}. For both the parallel and perpendicular
friction strong differences emerge with respect to the case of a
cubic box with periodic BC in all directions.

The structure of this paper is as follows: in \protect{\bf section~
\ref{ST}} the
simulation technique is introduced, with particular emphasis on the use
of the VP during the collision step; in \protect{\bf section \ref{PFTS}}, we
validate the model through the study of the Poiseuille flow in a slit,
for which the theory is known. In particular we calculate
the forces exerted by the solvent on the solid walls through two
independent methods, once by using the
VP and once using the stress tensor for a MPCD fluid near planar
boundaries. In \protect{\bf section~\ref{FSB}}, we compute  the
friction on a single
sphere in bulk. Because of the VP, the local Brownian
friction is increased, and we provide an Enskog-like model to
predict such an effect. In \protect{\bf section \ref{FSIS} }the
colloidal particle is
confined in a slit, and we measure the friction as a function of the
lateral size of the walls, while keeping fixed the separation between
the walls.
 Final remarks and observations are in \protect{\bf section \ref{C}}.

\section{\label{ST}Simulation technique} 
In the present application, Multi-Particle Collision Dynamics (MPCD) is a hybrid simulation
scheme, in which a coarse-grained approach is used to describe the
solvent variables, while an atomistic description is adopted for the
solvent-solute and for the solute-solute interactions. 
The dynamics of the system is
made up of two steps: streaming and collision. In the streaming step,
the position and  velocity of each particle is propagated for a time
$\delta t$ by solving
 Newton's equations of motion. In the collision step the fluid is
 subdivided into cubic cells of side $a$. Then a stochastic
 rotation of the particles velocities, relative to the center of mass motion
 of the relevant cell, is performed according to the formula:
\begin{equation}
\boldsymbol{V}_i(t)=\boldsymbol{u}+\boldsymbol{\Omega}\{\boldsymbol{v}_i(t)-\boldsymbol{u}\},
\label{eq.col1}
\end{equation}
where $\boldsymbol{u}$ is the mean velocity of the particles within a cell
and $\boldsymbol{\Omega}$ is a matrix which
rotates velocities by a fixed angle (in this work $\alpha~=~\pi/2$) around a randomly
oriented axis.
Through the stochastic rotation of the velocities, the solvent
particles can efficiently exchange momentum without introducing direct forces
 between them during the streaming step. As the collision step conserves
 mass, linear momentum and energy, the correct hydrodynamic
behavior is obtained on the mesoscopic scale
\cite{malevanets99,malevanets00}, as long as a shifted-grid procedure
is included in order to enforce Galilean invariance \cite{ihle01}. 
When colloids are present, Newton's
equations of motion are solved also for them \cite{padding06}.
Special care must
be used if no-slip boundary conditions (BC) are applied
on the colloidal surface and on the confining walls. In this
case, in fact,  we must couple  colloids and walls to the solvent 
during the collision step too. This is achieved by means of virtual
particles (VP). The
implementation of the no-slip BC is described in the following.\\ 
\protect{\em{Streaming step:}} when a MPCD particle
crosses the colloid (or wall) surface, it is moved back to the
impact position. Then a new velocity is extracted from the
following distributions for the tangent ($v_t$) and the normal
components ($v_n$) of the velocity, with respect to the surface velocity:
\begin{equation}
p(v_n)=\frac{m v_n}{k_BT}exp\left(-mv_n^2/2k_BT\right),\label{eq.vn}
\end{equation}
\begin{equation}
p(v_t)=\sqrt{m/2\pi k_BT}exp\left(-mv_t^2/2k_BT\right)\label{eq.vt},
\end{equation}
where $m$ is the mass of the solvent particle, $k_B$ Boltzmann's constant, and $T$ the
temperature of the system.
Once the
velocity has been updated, the particle is displaced for the remaining
part of the integration time step.\\
\protect{\em{Collision step:}} VP are inserted randomly
  in those parts of the system which are physically occupied by the
  colloid or by the walls (in sufficiently thick layers behind the interfacial positions).
  The VP density matches the MPCD solvent density $\gamma$,
  while their velocities $\boldsymbol{v}_i^{VP}$ are obtained from a Maxwell-Boltzmann
  distribution, whose average velocity is equal to the velocity of the
  colloid surface or to the velocity of the walls, and the temperature is the
  same as in the solvent. According to their coordinates, VP are
  sorted into the grid
cells.
 During the collision step, the average velocity of
  the center of mass of the cell is computed as
\begin{equation}
\boldsymbol{u}=\frac{
              \overset{n_{MPCD}}{\sum_{i=1}}\boldsymbol{v}_i(t)+
              \overset{n_{VP}}{\sum_{i=1}}\boldsymbol{v}_i^{VP}(t)
}{n_{MPCD}+n_{VP}},
\label{eq.newv}
\end{equation}
 where $n_{MPCD}$ and $n_{VP}$ respectively are the number of MPCD particles and the number of
 VP belonging to the same cell.
Finally velocities of both MPCD and VP
  belonging to
  the same cell are rotated according to the rule given in
  Eq.(\ref{eq.col1}). 

Due to the exchange of momentum between the solvent and the colloidal particle,
 the force exerted upon the latter may be expressed as:
 $\boldsymbol{F}=\boldsymbol{f}_s+\boldsymbol{f}_c$, where
 $\boldsymbol{f}_s$ and 
 $\boldsymbol{f}_c$ are the forces during the streaming step and the collision step, respectively. The former can be calculated as: 
\begin{equation}
\protect{\bf{f}}_{s}=-\frac{1}{\delta t}\overset{Q}{\sum_{i=1}}\Delta \protect{\bf{P}}_i^{MPCD},
\label{eq.forces}
\end{equation}
where $Q$ is the number of MPCD particles which have
crossed  the surface of the colloid between two
collision steps and $\Delta \boldsymbol{P}^{MPCD}_i$ is the change of momentum
of the $i$-particle of the solvent, which has been scattered by the
colloid.
The force exerted during the collision step is:
\begin{equation}
\protect{\bf{f}}_{c}=\frac{1}{\delta t}\overset{q}{\sum_{i=1}}\Delta
\protect{\bf{P}}_i^{VP},
\label{eq.forcec}
\end{equation}
where $q$ is the total number of virtual particles which belong to a
tagged colloid; and $\Delta \boldsymbol{P}^{VP}_i$ is the change of momentum of
the $i$- virtual particle during the collision step.
If walls with no-slip BC are present, we can also
measure the force exerted by the solvent upon them through
 Eqs.~(\ref{eq.forces}-\ref{eq.forcec}). In such a case $Q$ is the number of
solvent particles that have crossed the surface of a wall and $q$ is
the number of  VP belonging to the same wall.

\section{\label{PFTS}Poiseuille flow through a slit}
In this section we study the flow of the MPCD solvent through a slit
under the influence of an external uniform force
$\protect{\bf{F}}^{ext}$, which is oriented parallel to the walls.
 Walls are placed at $x=0$ and $x=L_x$ and
the lateral sides of the walls are $L_y=L_z$. In our simulations we
choose the solvent density equal to $\gamma=5$, and the interval between two
collision steps equal to $\delta t =0.1 t_0$. The time is in units of
$t_0=a(m/k_BT)^{1/2}$, where $m$ is the mass of the solvent particle,
$k_B$ the Boltzmann constant and $T$ the temperature. Hereafter we
assume that $k_BT=1$ and $m=1$.
A stationary parabolic velocity profile is expected to form for an incompressible fluid.
\begin{figure}[tb]
\includegraphics[width=7cm,height=7cm,angle=270]{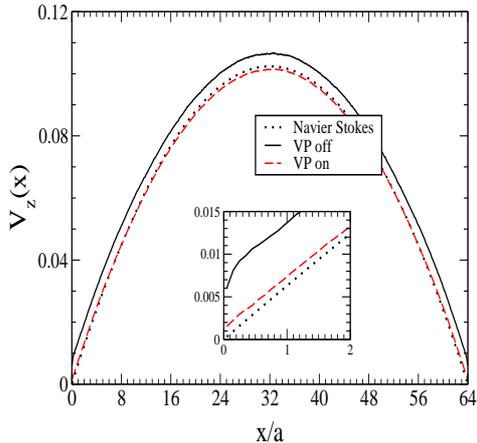}
\vspace{-0.5cm}
\caption{\label{fig1}\footnotesize{(Colors online) Velocity profile across
a slit of width $L_x$. Simulations are performed in a box of sides
$L_x=L_y=L_z=64a$.  The external applied field 
is $\boldsymbol{F}^{ext}=10^{-4}f_0\boldsymbol{\hat{z}}$. The inset
shows a magnification of the profile near the wall at $x=0$.}}
\end{figure}
Simulation results for the case $\protect{\bf{F}}^{ext}=10^{-4}f_0\boldsymbol{\hat{z}}$,
with $f_0=k_BT/a$
 are plotted in
 Fig.~(\ref{fig1}). When the no-slip BC are
 implemented only in the streaming step, using the stochastic
 reflections according to
 Eqs.~(\ref{eq.vn}-\ref{eq.vt}), the slope of the velocity profile (solid line) just
 near the wall is different from the expected one (dotted line); moreover 
 large slippage near the wall persists. When VP are included into the
 collision step, the $v_z(x)$ profile (dashed line) reproduces quite well the
 Navier-Stokes solution. The use of VP clearly ameliorates the
 simulation results with no-slip BC as expected\cite{lamura01,winkler09,gotze07}.
 We next ask what are the consequences of the various changes for the force exerted by the
 fluid upon the wall?  From a balance of external body and wall forces in a stationary state,
the expected value of the force on each wall
is $f^{exp}=\gamma F^{ext} \frac{L_x}{2}L_yL_z$ \cite{durst}, which in our
case is $f^{exp}=16f_0$. \\
In Fig.~(\ref{fig2}) we see that the total force converges to the expected value
 of $f^{exp}=16f_0$, at which point a stationary state is reached.
In the stationary state the contribution to the total force due to the collision step
 ($\boldsymbol{f}_c$) is relatively
 large. This happens especially when the mean free path of the MPCD fluid
 is small, in which case momentum is transported mainly via collisions
 rather than via diffusion.
\begin{figure}[tb]
\includegraphics[width=7cm,height=7cm,angle=270]{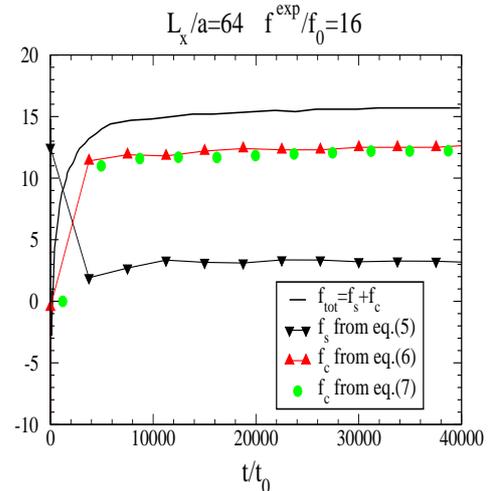}
\vspace{-0.5cm}
\caption{\label{fig2}\footnotesize{(colors online)Average force exerted by the
fluid upon each wall of a slit of width $L_x=64 a$, versus time. The
total expected value is $f^{exp}=16f_0$;  $f_s$ and $f_c$ are the force exerted
during the streaming and the collision steps. }}
\end{figure} 

For the slit geometry it is easy to adapt the general
expression for the stress tensor of the MPCD fluid  with periodic
BC \cite{ihle01,ihle03} in order to measure
the force just near the wall. The force exerted along the
$\boldsymbol{\hat{z}}$
direction because of collisional
 exchange of momentum along the $\boldsymbol{\hat{x}}$ direction
is:
\begin{equation}
f_c=\frac{m}{\delta t}\overset{p}{\sum_{j=1}}\Delta\xi_{j,x}^S\Delta
v_{j,z}.
\label{eq.shear} 
\end{equation}
In this expression we have used the same notation as adopted in
\cite{ihle01}, where a full description of the MPCD stress tensor
is provided. The meaning of the notation is as follows:
$\Delta\xi_{j,x}^S=\xi_{j,x}(t+\delta t)-\xi_{j,x}^S(t+\delta t)$,
$\Delta v_{j,z}=V_{j,z}(t+\delta t)-v_{j,z}(t+\delta t)$; $V_{j,z}$
and $v_{j,z}$ are the $z$ component
of the
velocities after and before the collision
step of particle $j$. $\xi_j(t)=(\xi_{j,x},\xi_{j,y},\xi_{j,z})$ are
the coordinates
of the cell of the unshifted grid containing the particle j at time
$t$. Similarly $\xi_j^S$ are the coordinates of the cell of the
shifted grid containing the $j$ particle at time $t$. The sum in
Eq.~(\ref{eq.shear}) applies only to the $p$ particles of the MPCD
solvent belonging
to the cells which overlap one of the walls.
As shown in Fig.~(\ref{fig2}), the expressions of the
forces according
to Eq.~(\ref{eq.forcec}) and (\ref{eq.shear}) are
 in very good 
agreement. However, if one needs the force exerted by the
fluid upon a spherical colloid or an irregularly shaped object, it
is important to take into account the local 
curvature of the colloidal surface. In such a case the generalization
 of the stress
tensor for the MPCD fluid is not easy, while the
implementation of Eq.~(\ref{eq.forcec}) is straightforward.

\section{\label{FSB}Friction on a sphere in the bulk}

We now turn to the translational friction on a sphere which is
not free to rotate. In this respect  we consider the case of a single
sphere of radius 
$R_{col}=4a$, which is kept at a fixed
position by applying a constraint force
$\boldsymbol{\mathcal{F}}^c$. This condition is similar to the case of
very massive particles embedded in solvent as studied in
\cite{whitmer10}. We use a
cubic simulation box whose side is $L=32a$. We are primarily interested in the
autocorrelation function of $\boldsymbol{\mathcal{F}}^c$
because this can be connected, via a Green-Kubo relation, to the translational
friction tensor \cite{akkermans00}:
\begin{equation}
\Xi_{\alpha\beta}=\lim_{t\rightarrow\infty}\frac{1}{k_BT}\int_0^{t}d\tau \left\langle \mathcal{F}_{\alpha}^c(t_0+\tau)\mathcal{F}_{\beta}^c(t_0)\right\rangle_{t_0}
\label{eq.friction}
\end{equation}
where $\alpha,\beta \in \{x,y,z\}$.

\begin{figure}[tb]
\includegraphics[width=7cm,height=8cm,angle=270]{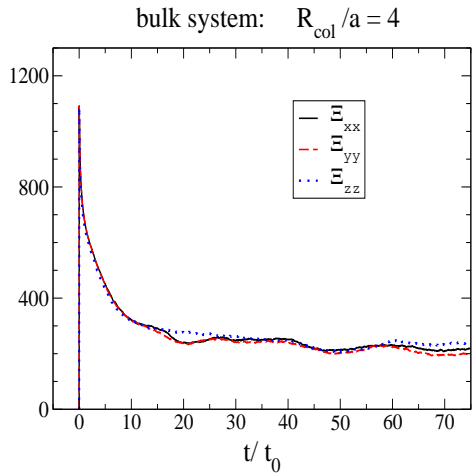}
\vspace{-0,5cm}
\caption{\label{fig3} \footnotesize{(Colors online) Running integral of the
autocorrelation functions of the constraint force. The off-diagonal terms of
Eq.~(\ref{eq.friction}), which are not shown here, go quickly to
zero. 
}}
\end{figure}

Because of symmetry, the friction does not depend on the cartesian direction along which
we measure, see Fig.~(\ref{fig3}).
If we consider a particular direction $\alpha$,
the long time
limit of $\Xi_{\alpha\beta}$ provides the total
friction $\xi$ upon the sphere, for which there are essentially two
contributions:
one coming from the local Brownian collisions with the particles
of the fluid ($\xi_E$), while the other is due to the long-range hydrodynamic
interactions ($\xi_S$). A simple empirical formula says that  the hydrodynamic
 and the Brownian friction should be added in parallel
 \cite{hynes77,lee04} in order to
 obtain the total friction:
\begin{equation} 
1/\xi= 1/\xi_E+1/\xi_S.
\label{vale}
\end{equation}
Both $\xi$ and $\xi_E$ may be calculated using our simulations. $\xi$ can be read from the long-time limit of the integral in Eq.(8),while
$\xi_E$ corresponds to the height of the short-time peak
\cite{padding10}. The
hydrodynamic term can be extracted by  inverting
Eq.(\ref{vale}). In the present case we obtain
$\xi_E^{sim}~=~1088~\pm~4$ and $\xi_S^{sim}~=~270~\pm~30$.\\
The hydrodynamic friction can be theoretically estimated from the  drag force
$\boldsymbol{F}_d$ on a colloid with no-slip BC in an infinite fluid
medium, according to the Stokes law
$\boldsymbol{F}_d=-\xi_S^{\infty}\boldsymbol{v}_{\infty}=-6\pi\eta
R_{col}\boldsymbol{v}_{\infty}$,
where $\boldsymbol{v}_{\infty}$ is the flow field at large
distances. As the Stokes friction depends on the long-range
hydrodynamic effects, it can be substantially affected by the
finite-size $L$ of the simulation box. Detailed calculations in
\cite{dunweg93} suggest that to lowest order in $R_{col}/L$, the
correction is given by $\xi_S=\xi_S^{\infty}\beta\left(R_{col}/L\right)$,
with 
\begin{equation}\beta\left(R_{col}/L\right)\approx
\left(1-2.837\left(R_{col}/L\right)\right)^{-1}
\label{eq:sizeeffects}
\end{equation}
Taking into account the finite-size effects, the value
 for the Stokes friction is $\xi_S= 291$, and the simulations result
 is in good 
agreement with this.

We now turn to the Brownian friction $\xi_E$, due to local collisions
with the solvent. A suitable starting point to calculate this term is the Enskog-Boltzmann
theory for a dense
gas. This model takes into account only two-bodies collisions and
 successive collisions are supposed to be uncorrelated. In other
words no influence of the local disturbance, induced by the tagged colloidal particle
on its surroundings, is included. The expression for the translational
friction in a bath of particles of
mass $m$ is the following
\cite{padding05}:
\begin{equation}
\xi'_E=\frac{8}{3}\left(\frac{2\pi k_BTmM}{m+M}\right)^{1/2}\gamma
R_{col}^2\frac{1+2\chi}{1+\chi},
\label{eq.enskog}
\end{equation}
where $\chi=I/MR_{col}^2=2/5$ is the gyration ratio for a
sphere. According to this model $\xi_E\sim700$, while from simulations
we obtain a much larger value. The discrepancy is
a consequence of the changes made to the collision step, as
the
exchange of momentum between the MPCD particles and the colloid, via
VP, changes the effective number of local Brownian collisions. 

In order to 
evaluate the
 fluctuations in the constraint force $\boldsymbol{\mathcal{F}}^c$ due to the VP,
 let us focus on one
collision cell which partly overlaps with a colloid  and contains $p$
MPCD particles and
$q$ VP, with velocities $\vv_i$ and $\ww_i$, respectively. Let us
suppose that the colloid moves with velocity $\VV$ through the ideal
gas bath, then we know that
\begin{eqnarray}
\left\langle \vv_i \right\rangle = \mathbf{0} & \qquad & \left\langle
v_i^2 \right\rangle = \frac{3k_BT}{m} \\
\left\langle \ww_i \right\rangle = \VV & \qquad & \left\langle
(\ww_i-\VV)^2 \right\rangle = \frac{3k_BT}{m}
\end{eqnarray}
The collision step itself may be written as (primes indicate
velocities after the collision):
\begin{eqnarray}
\boldsymbol{u} &=& \frac{1}{p+q} \left( \sum_{i=1}^{p} \vv_i + \sum_{i=1}^{q} \ww_i \right) \\
\vv_i' & =& \vv_i + (\boldsymbol{\Omega}-\II)  \left( \vv_i - \boldsymbol{u} \right) \\
\ww_i' & =& \ww_i + (\boldsymbol{\Omega} -\II)\left( \ww_i - \boldsymbol{u} \right)
\end{eqnarray}
where $\boldsymbol{\Omega}$ represents a rotation of an angle $\alpha$ around a
random axis, $\II$ is the unit matrix, and $\boldsymbol{u}$ the
average velocity of the cell center of mass.
The total change of momentum of the VP, i.e. their contribution to the change of momentum of the
colloid (or wall) is equal to:
\begin{eqnarray}
\Delta \PP^{VP} &=& m \sum_i^{q} \left( \ww_i' - \ww_i \right) \nonumber \\
&=& m (\boldsymbol{\Omega}-\II) \left[ \frac{p}{p+q} \sum_i^{q} \ww_i - \frac{q}{p+q} \sum_i^{p} \vv_i \right]. \nonumber \\
\end{eqnarray}
We define the force exerted on the colloid by the cell under consideration as
$\boldsymbol{F}^{VP}=\Delta\boldsymbol{P}^{VP} \delta t$ and
consider its contribution to the Enskog friction matrix:
\begin{eqnarray}
 \boldsymbol{\Xi}(t) &=& \beta \int_0^t \left\langle \FF(\tau) \FF(0) \right\rangle \dif \tau \nonumber \\ 
&=& \frac12 \beta \left\langle \FF(0) \FF(0) \right\rangle \delta t +
\sum_{j=1}^i \left\langle \FF(j \delta t) \FF(0) \right \rangle \delta t. \nonumber \\
\end{eqnarray}
Only the first term will be non-zero because velocities at different
times are uncorrelated. When calculating this term, we assume
$<\boldsymbol{v}_i\boldsymbol{v}_j>=\frac{k_BT}{m}\II\delta_{i,j}$,
$<\boldsymbol{w}_i\boldsymbol{w}_j>=\frac{k_BT}{m}\II\delta_{i,j}$ and
$<\boldsymbol{v}_i\boldsymbol{w}_j>=0$, which turns $\boldsymbol{\Xi}=\boldsymbol{\Xi}(0)$ into a diagonal matrix.
Averaging over all possible orientations of the rotation axis we obtain for the diagonal elements:
\begin{equation}
\xi_{VP} = \frac23 (1-\cos \alpha ) \frac{m}{\delta t} \frac{p q}{p + q}.
\end{equation} 
Since velocities in different cells are uncorrelated we may simply add the contributions of all
cells overlapping with the colloid, obtaining:

\begin{equation}
\xi_{VP}^{tot} = \frac23 (1-\cos \alpha) \frac{m}{\delta t_c} \sum_{\mathrm{cells}\ k} \frac{p_k q_k}{p_k+q_k}.\label{ventuno}
\end{equation}

Performing the sum over cells analytically is rather complicated, if not impossible, so we have decided to evaluate Eq.~(\ref{ventuno}) numerically, during the simulation
itself. Including the correction so obtained, the predicted short
time friction becomes:
$\xi_E~=~\xi'_E~+~\xi_{VP}^{tot}~=~1084$, which is in very good agreement
with the simulation value $\xi_E^{sim}~=~1088 ~\pm~ 4$.

\section{\label{FSIS}Friction on a sphere inside a slit}

Recent developments in the field of micro- and nanofluidics have led
to a renewed interest in molecular hydrodynamics phenomena near a
solid surface. It has been shown theoretically
\cite{ganatos80,happel91,cichocki98,bhattacharya05} and
experimentally \cite{malysa86,lobry96,lin00,leach09}
 that the hydrodynamic interactions with
flat walls can slow down the motion of colloidal particles
substantially,
and that such effect also
depends on the direction
of the motion of the particle (for example parallel or perpendicular to
the walls).
We therefore apply our method to study the friction on a
sphere inside a slit. 
 In this case, as in the bulk system, we do not take into
 account the role played by the angular momentum, so that the present
 discussion is relevant only for purely translating spheres in a slit.
The walls are set at $z=0$ 
and $z=L_z$ while we use periodic BC along the $x$ and $y$
directions.The length of the walls' sides is $L_x=L_y=L$.
We have chosen this particular configuration because we
have access to analytical solutions of the Navier-Stokes equations for
the parallel motion \cite{happel91} and numerical solutions for the
perpendicular motion as well \cite{bhattacharya05}. Such solutions
have been provided under the assumption that the lateral size ($L$) of
the walls
 is infinitely large.

Our aim is to study how the friction
depends on the lateral width $L$ of the cell when the walls separation $L_z$
is fixed. Besides establishing finite size effects for the usual simulation boxes, this topic is also relevant for the study of experimental periodic
arrangements of particles, such as trains and grids of particles under
confinement. In both cases a periodic array of colloidal particles exists, exerting strong hydrodynamic interactions on each other.\\
Simulations have been performed keeping the colloidal particle of Sec.\ref{FSB}
 in a fixed
position in the mid-plane of the slit. 
From the autocorrelation of the constraint force we obtain
the friction coefficients (as explained in section
\ref{FSB}) in the directions parallel and perpendicular
to the walls. 
Under the assumption of small Reynolds numbers,
it is possible to express the effects of the walls by means of a correction factor
 $\tau$
 to the Stokes friction $\xi_S^{\infty}$ in an unbounded system:
\begin{equation}
\xi^{\perp}_1=\tau^{\perp}\xi_S^{\infty}.
\label{eq:stokesslit}
\end{equation}
Similarly we describe finite-size effects by a second correction factor $\beta$:
\begin{equation}
\xi^{\perp}_2=\beta^{\perp}\tau^{\perp}\xi_S^{\infty}.
\label{eq:stokesslitfinita}
\end{equation}
In the following we will obtain $\beta^{\perp}$ as
\begin{equation}
\beta^{\perp}=\xi^{\perp}_{sim}/\xi_1^{\perp}.
\end{equation}
Similar expressions hold for the parallel frictions. 

Analytical expressions for the parallel and perpendicular
frictions in a slit geometry are not available, but analytical expression for
a single wall do exist \cite{happel91}.
The correction for the parallel friction due to the presence of a single wall is:
\begin{eqnarray}
\tau^{||}&=&\left[ 1-1.004(R_{col}/z)+0.418(R_{col}/z)^3+\right. \nonumber \\
&& \left.+0.21(R_{col}/z)^4-0.169(R_{col}/z)^5\right]^{-1},
\label{eq:faxenslit}
\end{eqnarray}
whereas the correction for the perpendicular friction is: 
\begin{eqnarray}
\tau^{\perp}&=&
\frac{4}{3}\sinh(\alpha)\sum_{_n=1}^{\infty}\frac{n(n+1}{(2n-1)(2n+3)}\times \nonumber \\
&& \left(
\frac{2\sinh(2n+1)\alpha+(2n+1)\sinh2\alpha}{4\sinh^2(n+0.5)\alpha-(2n+1)^2\sinh^2(\alpha)}-1
\right), \nonumber \\
\label{eq:perpendicular}
\end{eqnarray}
where $\alpha=\cosh^{-1}\left(z/R_{col}\right)$ and $z$ is the distance
of the particle from the wall. As a first approximation, we use linear
superposition theory, according to which we can add the contributions
coming from each wall as if they behave independently from one another:
\begin{equation}
\tau_{\infty}^{\perp}\approx \tau^{\perp}_{w1}+
\tau^{\perp}_{w2}-1,
\label{eq:super}
\end{equation}
the subscripts $w1$ and $w2$ refer to the two walls. We have subtracted 1, in order to avoid counting the bulk contribution to the
friction twice. This is a mere consequence of the definition adopted for the
corrective terms. 

In case the walls are very far apart, this approximation works fairly well,
otherwise it tends to overestimate the combined effects of the two
walls. 
In particular, for the simulation parameters we have used here
($R_{col}=4a$ and $L_z=32a$), the
friction obtained with the superposition model appears to be overestimated
by about $15\%$ with respect to the numerical solution of the
Navier-Stokes equations in the slit as given in
\cite{bhattacharya05}. In the rest
of this section we will use the results provided by
Eq.(\ref{eq:super}) but corrected according to
\cite{bhattacharya05}.

\begin{figure}[tb]
\includegraphics[width=6cm,height=8cm,angle=270]{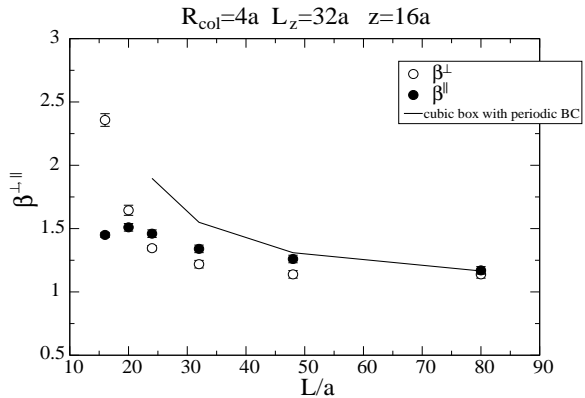}
\caption{\label{fig6} \footnotesize{ Finite-size
    effects corrections for the friction upon a spherical particle
    placed in the mid-plane of a slit geometry. The lateral size of
    the square walls is $L$, while the wall separation is
    $L_z=32a$. Circles: simulation results. Solid line:
    the corrective terms for a cubic box with periodic BC in all three
    directions (Eq.\ref{eq:sizeeffects}).
}}
\end{figure}

The values of $\beta^{\perp,||}$ are
plotted in Fig.~(\ref{fig6}) as a function of lateral size $L$.

For values of $L<24a$, the
behavior of $\beta^{||}$ and $\beta^{\perp}$ is very different: with decreasing
$L$, $\beta^{||}$ decreases while $\beta^{\perp}$ increases.
For the parallel motion we expect that, when the lateral size $L$
is very small and the particle is very close to its
images, a wake effect is established which reduces the
friction along the lines connecting the colloid and its nearest images. A
similar trend can be deduced from
\cite{bhattacharya08,kohale08}. The situation depicted in these works
is slightly different from ours. They for example study the parallel
friction as a function of $L_x$ once $L_y$ and $L_z$ are fixed, while
in the present simulations we always use a square periodic space
($L_x=L_y$). However if, for each profile shown in
\cite{bhattacharya08,kohale08}, we pick up the points corresponding to
$L_x=L_y$, it is seen that the friction decreases as
$L$ diminishes. On the contrary when $L>24a$, the parallel and perpendicular
frictions behave similarly: they decay slowly to the case of a slit
whose walls are infinitely large. It is possible that the screening
point where we
can observe different behaviors between the parallel and
perpendicular friction is for $L\sim L_z$.

When $L=L_z$, the simulation box is cubic, so we can compare our
results with those provided by the correction factors in 
  Eq.~(\ref{eq:sizeeffects})  plotted in 
 Fig.~(\ref{fig6}) as a solid line.
We see that there exists a
significant difference which is due just to the presence of the
confining walls: the latter screen even the effects due to the use of
periodic BC. 
However, once the wall separation $L_z$ is fixed, the correction factor decays
  slower when compared to Eq.~(\ref{eq:sizeeffects})
    where periodic BC are used in all directions (Fig.~(\ref{fig6})).

\begin{figure}[tb]
\includegraphics[width=6cm,height=8cm,angle=270]{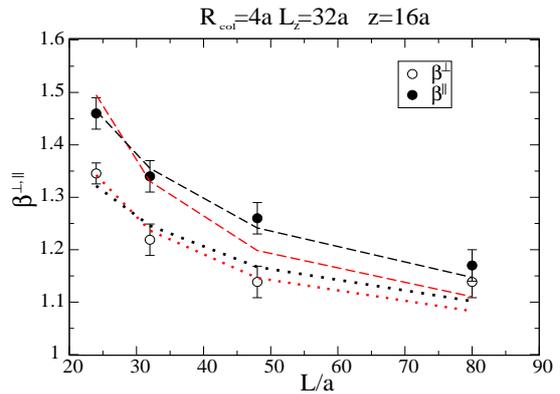}
\caption{\label{fig7} \footnotesize{(Colors online) Finite-size
    effects for $L>20$ in comparison with two different fit
    functions. Red lines
    are based on a function of the type as in 
      Eq.~(\ref{eq:sizeeffects}); black lines are based upon the
    function in Eq.~(\ref{eq:log}). Dotted lines are for
    the perpendicular friction, dashed lines for the parallel one.
}}
\end{figure}

For ease of use, we have tried to fit the data for $L>20$ with a
function similar to that in Eq.~(\ref{eq:sizeeffects}) (see red lines in Fig.~(\ref{fig7})):
the behavior seems good for the perpendicular friction but it
deteriorates for the parallel friction.
We have also tried a fit with the
function 
\begin{equation}
f(L)=1+A\ln(1+R_{col}/L),\label{eq:log}
\end{equation}
 in which $A$
is a parameter whose value depends on the direction we consider: for
the parallel friction $A^{||}~\sim ~2.08$ while for the perpendicular
 motion $A^{\perp}~\sim 3.02$. Eq.~(\ref{eq:log}) seems to work better
 than Eq.~(\ref{eq:sizeeffects}). For example the parallel friction is
 better captured
 (see black lines in Fig.~(\ref{fig7})).
We do not have, however,
indications whether the parameter $A$ is a function of
the wall separation $L_z$ and of the particle position $z$. To clarify this
point further simulations are necessary, to be compared with the
solutions of the Navier-Stokes equations for the slit for different
particle positions.

In summary, our results suggest that:
\begin{itemize}
\item There are two different regimes depending on whether the lateral
  size of the walls is smaller or larger than the wall separation.
\item For a cubic box, the finite size effects in the slit geometry
  are less than in a system with periodic BC in all three directions.
\item Once $L_z$ is fixed and $L>L_z$, the correction factors
  $\beta^{\perp,||}$
  decay slowly towards unity, possibly as a logarithmic function of $L$.
\end{itemize}

\section{\label{C}Conclusions}

In conclusion we have described a protocol for the MPCD simulation
technique, which provides a satisfactory treatment of the no-slip
boundary conditions (BC) and the correct evaluation of the force
exerted by the solvent on solid surfaces. Such a protocol is based
upon the use of virtual particles during the collision-step, a method
already used in literature and which appears suited to study
wall-liquid interfaces and very massive particles embedded in the
solvent, or particles kept fixed by an external constraint force as in
the present work. We pay particular attention to the calculation of
the friction tensor for a non rotating sphere.\\
We have shown
how the no-slip BC modifies the local Brownian friction on a spherical
colloid, and how such effects can be evaluated through an
Enskog-like treatment of the VP. We have also studied the friction on
a particle embedded in a slit, analyzing the dependence of the
simulation results on the use of periodic boundary conditions along
the sides of the walls. When the lateral size of the walls is very
small (less than the wall separation), strong coupling effects between
a particle and its images are observed. Moreover, the parallel and the
perpendicular friction show an antithetic behavior: as $L$ decreases,
the parallel friction decreases too, while the perpendicular friction
increases. When $L>L_z$, the corrective terms for the parallel and
perpendicular friction are much more similar to each other and they
both decrease to unity as $L$ increases, approaching more and more the
case of an ideal slit made of two infinitely large walls. 
 When $L=L_z$, we find that the finite-size effects (FSE) in the slit
 geometry are less than for a cubic box with periodic boundary
 conditions in all directions. Moreover, once $L_z$ is fixed and
 $L>L_z$, the FSE in the slit appear to be slowly
 varying with the system size $L$, possibly according to a logarithmic
 function of $R_{col}/L$.\\
Further studies will concern whether and how the FSE in a slit depend
on the walls' separation $L_z$ and on the particle distance $z$. 

\bibliography{paper1}
\end{document}